# INERTIAL DYNAMICS AND EQUILIBRIUM CORRELATION FUNCTIONS OF MAGNETIZATION AT SHORT TIMES

Sergei V. Titov[1], William J. Dowling[2], Anton S. Titov[3], Sergey A. Nikitov[4], and Mikhail Cherkasskii[5,*, ‡]

[1]*Kotel'nikov Institute of Radioengineering and Electronics of the Russian Academy of Sciences, Fryazino, Moscow Region 141190, Russia*

[2]*Department of Electronic and Electrical Engineering, Trinity College Dublin, Dublin 2, Ireland*

[3]*The Moscow Institute of Physics and Technology (State University), Institutskiy per. 9, Dolgoprudnyy, Moscow Region, 141701, Russia*

[4]*Kotel'nikov Institute of Radioengineering and Electronics of the Russian Academy of Sciences, Moscow 125009, Russia*

[5]*Faculty of Physics, University of Duisburg-Essen, Duisburg 47057, Germany*

**Abstract**

The method of moments is developed and employed to analyze the equilibrium correlation functions of the magnetization of ferromagnetic nanoparticles in the case of inertial magnetization dynamics. The method is based on the Taylor series expansion of the correlation functions and the estimation of the expansion coefficients. This method significantly reduces the complexity of analysis of equilibrium correlation functions. Analytical expressions are derived for the first three coefficients for the longitudinal and transverse correlation functions for the uniaxial magnetocrystalline anisotropy of ferromagnetic nanoparticles with a longitudinal magnetic field. The limiting cases of very strong and negligibly weak external longitudinal fields are considered. The Gordon sum rule for inertial magnetization dynamics is discussed. In addition, we show that finite analytic series can be used as a simple and satisfactory approximation for the numerical calculation of correlation functions at short times.

* Corresponding author: macherkasskii@hotmail.com

‡ Present address: Institute for Theoretical Solid State Physics, RWTH Aachen University, 52074 Aachen, Germany

## 1. Introduction

Spin dynamics is of fundamental importance in spintronics and magnonics research. To incorporate ultrafast phenomena, more accurate mathematical models of the spin dynamics must be developed. The prominent Landau-Lifshitz-Gilbert (LLG) equation, which adequately addresses picosecond spin precession, has been generalized to consider spin inertia manifesting as sub-picosecond nutation [1–23]. Although significant progress has been achieved in the study of ultrafast phenomena, the study of spin inertia has only just begun.

The inertial spin effects were introduced in the LLG equation by mesoscopic nonequilibrium thermodynamics theory [3], using the torque-torque correlation model [12], as memory effects [4], with the classical Lagrangian approach [5,16,19] and by Dirac quantum theory [14]. As a result of these investigations, the term with the second-order time derivative of magnetization was included in the LLG equation, which is then referred to as the inertial Landau-Lifshitz-Gilbert (ILLG) equation. This equation reads

$$\dot{\mathbf{u}} = \gamma[\mathbf{H}\times\mathbf{u}] + \alpha[\mathbf{u}\times\dot{\mathbf{u}}] + \tau[\mathbf{u}\times\ddot{\mathbf{u}}], \tag{1}$$

where $\mathbf{u} = \mathbf{M}/M_S$ is a unit vector along the magnetization **M**, $M_S = |\mathbf{M}|$ is the saturation magnetisation, $\mathbf{H}$ is the effective magnetic field, $\alpha$ is the Gilbert precession damping coefficient, $\gamma = 2.2\times 10^5\ \mathrm{rad\cdot m\cdot A^{-1}\cdot s^{-1}}$ is the gyromagnetic ratio and $\tau$ is the inertial relaxation time. The ILLG equation considers spin precession caused by Zeeman interaction, Gilbert damping, which can be attributed to first-order relativistic spin-orbit coupling, and inertia of spin originating due to second-order relativistic spin-orbit effect [13,14]. The last term in Eq. (1) is reminiscent of the dynamics of a massive particle, and causes nutation of magnetization superimposing on the precession. Nutation resonance was recently detected at terahertz frequencies in NiFe, CoFeB and Co ferromagnetic films [1,2] at a frequency, which is proportional to $\tau^{-1}$. In addition, it was found that inertia induces red-shift of the ferromagnetic resonance [7,17,20]. The non-inertial LLG equation, used successfully at gigahertz frequencies, predicts the divergence of the integral spectral power, or in other words, a plateau of the absorption coefficient at ultrahigh frequencies, which is equivalent to the notorious plateau in the non-inertial Debye model of dielectric relaxation. Adding the inertial term to the LLG equation solves this problem.

Theoretical investigation of magnetization trajectories within the ILLG model demonstrates the relatively complex behavior, which should be considered in the interpretation of the experimental results. For instance, the solutions for the non-damped limit of the ILLG equation were obtained in terms of the Jacobi elliptic functions and elliptic integrals for a single-domain ferromagnetic nanoparticle [18]. Thus, the task of finding the analytical solution of the ILLG equation to calculate the correlation functions and the susceptibility is a formidable one. For the damped case, numerical simulations were presented in Refs. [7,10,21]. However, such numerical studies provide an abundance of information, which is difficult to interpret directly.

Here we present the method of moments employed for the analysis of the equilibrium correlation functions and the magnetic susceptibility of ferromagnetic nanoparticles demonstrating inertial magnetization dynamics. The method is based on the Taylor series expansion of the correlation functions of magnetization about the initial time, $t=0$, and the analytical estimation of the expansion coefficients (moments), namely [24–30]

$$S_n = (-i)^n \frac{d^n}{dt^n} C(t)\bigg|_{t=0} . \tag{2}$$

According to the definition in Eq. (2) the moments are determined by the short time behavior of the correlation function $C(t)$. Therefore, the moments imply information about the shape of the correlation function and particularly about its behavior on a short time interval. According to linear response theory, there is a relationship between the susceptibility and the equilibrium correlation function [26]. Consequently, the method of moments can be effectively used to analyze the complex shape of spectra.

The method of moments enables the analysis of spectra in terms of equilibrium magnetization parameters (inertial relaxation time, temperature, direction cosines of the magnetization vector) and the parameters of the free energy. In this method it is not required to solve the ILLG equation and to find time-dependent correlation functions. Thus, the determination of the spectral moments is a simpler problem than the calculation of spectral band contours. As a result, a natural interpretation of the magnetic susceptibility spectra can be obtained.

## 2. Correlation functions for uniaxial magnetocrystalline anisotropy with a longitudinal magnetic field

Most ferromagnetic solids are magnetically anisotropic. Here we consider crystal systems having a single axis of high symmetry (easy axis). The magnetocrystalline anisotropy of such crystals is called uniaxial anisotropy [31,32]. The total dimensionless free energy of a uniaxial nanomagnet subject to a uniform d.c. magnetic field $\mathbf{H} = H\mathbf{e}_Z$ applied along the easy axis is [31]

$$\frac{V(\vartheta,\varphi)}{kT} = -\sigma\cos^2\vartheta - \xi\cos\vartheta\,, \tag{3}$$

where $\sigma = vK_u/(kT)$ is the dimensionless anisotropy parameter, $K_u$ is the uniaxial anisotropy constant, $\xi = v\mu_0 M_S H/(kT)$ is the external d.c. magnetic field parameter, $v$ denotes the volume of the nanomagnet, $kT$ is the thermal energy, and $\mu_0 = 4\pi\times 10^{-7}\,\mathrm{JA^{-2}m^{-1}}$ is the permeability of free space in SI units. The uniaxial anisotropy is commonly used in treatments of the LLG equation [33–38] (Fig. 1).

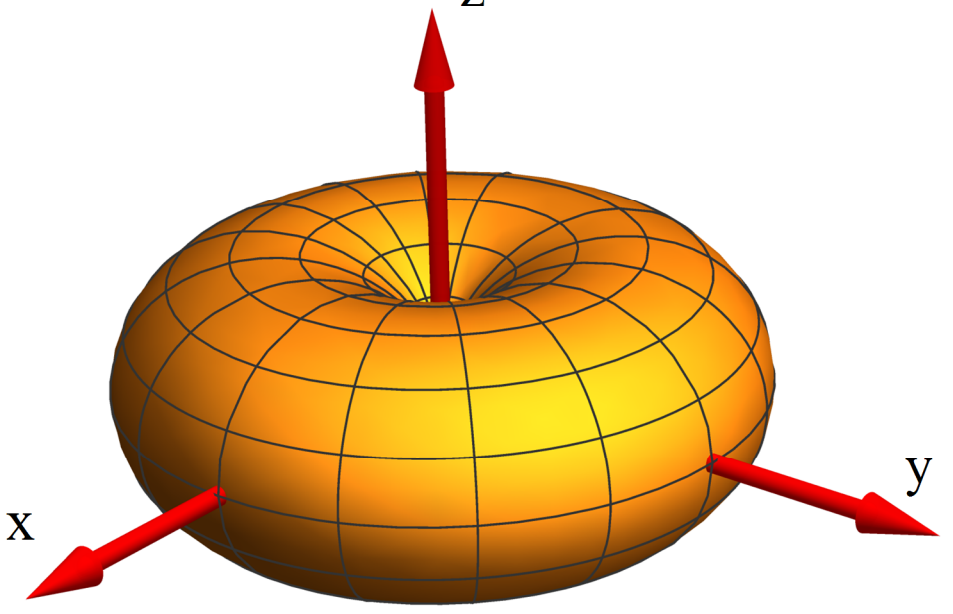

FIG. 1. Three-dimensional plot of the free energy $V(\vartheta,\varphi)$, Eq. (3), for $\sigma = 0$ and $\xi = 0.7$.

The magnetic susceptibility of anisotropic materials is a matrix known as the susceptibility tensor. Here we consider two different components of this complex tensor, namely, longitudinal (along the easy axis) susceptibility $\chi_\parallel$ and the transverse (perpendicular to easy axis) one $\chi_\perp$ for uniaxial magnetocrystalline anisotropy with a longitudinal magnetic field

$$\chi_\gamma(\omega) = \chi'_\gamma(\omega) - i\chi''_\gamma(\omega) = \int_0^\infty \chi_\gamma(t)e^{-i\omega t}dt\,, \quad \gamma = \parallel, \perp\,. \tag{4}$$

According to linear response theory and the fluctuation-dissipation theorem [26], the susceptibility $\chi_\gamma$ and the equilibrium correlation function $C_\gamma(t)$ are connected via the following expression

$$\chi_\gamma(\omega) = C_\gamma(0) - i\omega\tilde{C}_\gamma(\omega)\,, \tag{5}$$

where $\tilde{C}_\gamma(\omega)$ is the Fourier-Laplace transform of $C_\gamma(t)$

$$\tilde{C}_\gamma(\omega) = \int_0^\infty C_\gamma(t) e^{-i\omega t} dt. \tag{6}$$

The temporal correlation functions are used for the description of the temporal evolution of microscopic variables, such as the components of magnetization of fine ferromagnetic particles, and its influence on the value of the same microscopic variables later. These correlations are important in equilibrium systems, because a time-invariant macroscopic ensemble can still have non-trivial temporal dynamics microscopically.

Next, we analyze the longitudinal correlation function $C_\parallel(t)$ for the longitudinal component of magnetization $u_\parallel(t) = M_\parallel(t)/M_S$ and the transverse correlation function $C_\perp(t)$ for the transverse component of magnetization $u_\perp(t) = M_\perp(t)/M_S$. These correlation functions are defined as

$$C_\parallel(t) = \langle u_\parallel(0) u_\parallel(t) \rangle - \langle u_\parallel(0) \rangle^2, \tag{7}$$

and

$$C_\perp(t) = \langle u_\perp(0) u_\perp(t) \rangle, \tag{8}$$

where $u_\parallel(t) = \cos\vartheta(t)$ and $u_\perp(t) = \sin\vartheta(t)\cos\varphi(t)$ are the longitudinal and transverse components of a unit vector $\mathbf{u} = (\sin\vartheta\cos\varphi, \sin\vartheta\sin\varphi, \cos\vartheta)$ along the magnetization **M**. The brackets $\langle\ \rangle$ denote the equilibrium ensemble averages in the four-dimensional phase space $\left\{\vartheta_0 = \vartheta(0), \varphi_0 = \varphi(0), \omega_x^0 = \dot{\vartheta}\big|_{t=0}, \omega_y^0 = \dot{\varphi}\sin\vartheta\big|_{t=0}\right\}$ [39]

$$\langle(\cdot)\rangle = \int_0^\pi \int_0^{2\pi} \int_{-\infty}^{\infty} \int_{-\infty}^{\infty} (\cdot) W_{st}(\vartheta_0, \varphi_0, \omega_x^0, \omega_y^0) \sin\vartheta_0 \, d\omega_x^0 \, d\omega_y^0 \, d\varphi_0 \, d\vartheta_0, \tag{9}$$

where

$$W_{st}(\vartheta_0, \varphi_0, \omega_x^0, \omega_y^0) = Z^{-1} e^{-(\eta\omega_x^0)^2 - (\eta\omega_y^0)^2 - V(\vartheta_0,\varphi_0)/kT} \tag{10}$$

is the stationary distribution in the spin system [19],

$$Z = \int_0^\pi \int_0^{2\pi} \int_{-\infty}^{\infty} \int_{-\infty}^{\infty} \exp\left[-(\eta\omega_x^0)^2 - (\eta\omega_y^0)^2 - \frac{V(\vartheta_0,\varphi_0)}{kT}\right] \sin\vartheta_0 \, d\omega_x^0 \, d\omega_y^0 \, d\varphi_0 \, d\vartheta_0 \tag{11}$$

is the partition function, $V(\vartheta_0, \varphi_0)$ is the free energy given by Eq. (3), and the inertial parameter is defined as [19]

$$\eta = \sqrt{\frac{v\mu_0 M_S \tau}{2\gamma kT}}\,. \tag{12}$$

The stationary solution, Eq. (10), plays an important role in various applications such as escape rate theory [39]. The partition function *Z*, Eq. (11), can be calculated analytically and is given by

$$Z = \frac{2\pi^2 e^{\sigma}}{\eta^2 \sigma} B(\xi,\sigma)\,, \tag{13}$$

where

$$B(\xi,\sigma) = \sqrt{\sigma}\left[ e^{\xi} F_D\left(\frac{\xi + 2\sigma}{2\sqrt{\sigma}}\right) - e^{-\xi} F_D\left(\frac{\xi - 2\sigma}{2\sqrt{\sigma}}\right)\right] \tag{14}$$

and

$$F_D(x) = e^{-x^2}\int_0^x e^{y^2}\,dy \tag{15}$$

is the Dawson integral.

The above correlation functions have been defined in the context of equilibrium statistical mechanics, and therefore the Boltzmann distribution is used in the calculations. The important feature of the stationary Boltzmann distribution is that it contains the integral of motion for undamped motion under the exponent. Hence the correlation functions are calculated by the average of trajectories of magnetization over the initial values of the phase space variables (the equilibrium distribution does not change in time). In the other words we average over all possible trajectories, which are determined by the initial conditions.

The behaviour of the correlation functions $C_g(t)$ defined by Eqs. (7) and (8) is completely determined by the functions $u_\parallel(t)$ and $u_\perp(t)$, which must be found from the solution of the ILLG equation (1) with the effective field **H** expressed as [40]

$$\mathbf{H} = -\frac{1}{v\mu_0 M_S}\frac{\partial V}{\partial \mathbf{u}}\,. \tag{16}$$

### 3. Moments for longitudinal and transverse correlation functions

The correlation function $C_g(t)$ can be expanded in the Taylor series with the coefficients, which are related to the time derivatives of the correlation function $C_g(t)$ at $t = 0$ [24,25,27]

$$C_g(t) = S_0^g + \sum_{n=1}^{\infty} S_n^g \frac{(it)^n}{n!}, \tag{17}$$

where $S_n^g = (-i)^n \frac{d^n}{dt^n} C_g(t)\Big|_{t=0}$. For realistic functions of free energy, the coefficients of the power series expansion are finite, so the series converges over some range of time. The evaluation of the moments $S_n^g$ for magnetization correlation functions requires only the calculation of the time derivatives at $t = 0$ of the functions $u_{\pi}(t)$ and $u_{\perp}(t)$, viz.,

$$S_n^g = (-i)^n \left( \left\langle u_g(0) \frac{d^n u_g(t)}{dt^n} \right\rangle \Big|_{t=0} - \left\langle u_g(0) \right\rangle^2 \delta_{n0} \right). \tag{18}$$

Therefore the calculation of moments is reduced to the calculation of multiple integrals (see Eq. (9)), while both the averaged functions and the exponential distribution function are determined by the coordinates and velocities taken at time $t = 0$ . Thus there is no need to solve the ILLG equation (1) for finding the moments.

At the high-frequency limit ($\omega \to \infty$), the spectrum of the correlation functions $C_g(t)$, which are real functions of time, can be expanded as (see Eqs. (6) and (17))

$$\tilde{C}_g(\omega) = -\sum_{n=0}^{\infty} \frac{iS_n^g}{\omega^{n+1}}. \tag{19}$$

Moreover, using Eqs. (5) and (19), the spectrum $\chi_g(\omega)$ can be expanded in powers of $\omega^{-1}$

$$\chi_g(\omega) = -\sum_{n=1}^{\infty} \frac{S_n^g}{\omega^n}. \tag{20}$$

Both the correlation functions and the susceptibility can be analyzed using the moments. However, in practical calculations it is difficult to determine all the moments, so the correlation function and susceptibility are approximated by a finite series (the summation index in Eqs. (17) and (20) $n \le n_{\max}$). Thus, the series approximation with a finite number of terms corresponds to the correlation function only at short times Eq. (17), and corresponds to the susceptibility only at high frequencies Eq. (20).

## 4. Analytical expressions for the moments in the case of uniaxial anisotropy with a longitudinal magnetic field

Here we evaluate the moments $S_n^\gamma$ for the circular symmetric free energy Eq. (3). The similarity between the inertial magnetization dynamics and the rotation of a rigid body (like a symmetric top) [19] imposes a restriction on the correlation functions. The time correlation functions of classical ensembles of rigid molecules are even functions of time [27]. This restriction means that only the even moments $S_{2n}^\gamma$ are non-zero.

Equation (18) gives for the moments $S_0^\parallel = \langle \cos^2\vartheta_0 \rangle - \langle \cos\vartheta_0 \rangle^2$ and $S_0^\perp = \langle \sin^2\vartheta_0 \cos^2\varphi_0 \rangle$ after performing the ensemble average

$$S_0^\parallel = -\frac{1}{2\sigma} + \frac{\cosh\xi + \frac{\xi}{2\sigma}\sinh\xi}{B(\xi,\sigma)} - \frac{\sinh^2\xi}{B^2(\xi,\sigma)}, \tag{21}$$

$$S_0^\perp = \frac{1}{2}\left[1 - \left(\frac{\xi}{2\sigma}\right)^2 + \frac{1}{2\sigma} - \frac{\cosh\xi - \frac{\xi}{2\sigma}\sin\xi}{B(\xi,\sigma)}\right]. \tag{22}$$

The higher the order of the moment, the more complicated the analytical expression for the moment becomes. Next, for simplicity, we calculate the moments $S_{2n}^\gamma$ for $n = 1$ and 2 only. Noting that

$$\dot{u}_\parallel(0) = -\omega_x^0 \sin\vartheta_0, \tag{23}$$

$$\dot{u}_\perp(0) = \omega_x^0 \cos\vartheta_0 \cos\varphi_0 - \omega_y^0 \sin\varphi_0, \tag{24}$$

and performing the ensemble average $S_2^\gamma = -\langle u_\gamma(0)\ddot{u}_\gamma(0) \rangle = \langle \dot{u}_\gamma^2(0) \rangle$, we obtain the second-order moments

$$S_2^\parallel = \frac{1}{2\eta^2}\left[1 - \left(\frac{\xi}{2\sigma}\right)^2 + \frac{1}{2\sigma} - \frac{\cosh\xi - \frac{\xi}{2\sigma}\sinh\xi}{B(\xi,\sigma)}\right], \tag{25}$$

$$S_2^\perp = \frac{1}{4\eta^2}\left[1 - \frac{1}{2\sigma} + \left(\frac{\xi}{2\sigma}\right)^2 + \frac{\cosh\xi - \frac{\xi}{2\sigma}\sin\xi}{B(\xi,\sigma)}\right]. \tag{26}$$

The weighted sum of the moments $S_2^\parallel$ and $S_2^\perp$ is given by

$$\frac{1}{3}S_2^\parallel + \frac{2}{3}S_2^\perp = \frac{1}{3\eta^2}. \tag{27}$$

The sum rule in Eq. (27) does not depend on the anisotropy $\sigma$ and field $\xi$ parameters and coincides with the sum rule for isotropic media [28].

To evaluate the moments $S_4^{\parallel}$ and $S_4^{\perp}$ we need the time derivatives $\ddot{u}_{\parallel}(0)$ and $\ddot{u}_{\perp}(0)$ (see Eq. (18)), which depend on the time derivative of the angular velocity components, $\omega_x$ and $\omega_y$, viz.,

$$\ddot{u}_{\parallel}(0) = -\dot{\omega}_x^0 \sin\vartheta_0 - \left(\omega_x^0\right)^2 \cos\vartheta_0, \tag{28}$$

$$\begin{aligned}\ddot{u}_{\perp}(0) = {} & \dot{\omega}_x^0 \cos\vartheta_0 \cos\varphi_0 - \left(\omega_x^0\right)^2 \sin\vartheta_0 \cos\varphi_0 \\ & - \omega_x^0 \omega_y^0 \cot\vartheta_0 \sin\varphi_0 - \dot{\omega}_y^0 \sin\varphi_0 - \left(\omega_y^0\right)^2 \csc\vartheta_0 \cos\varphi_0 .\end{aligned} \tag{29}$$

The derivation of the expressions for $\dot{\omega}_x$ and $\dot{\omega}_y$ is given in Appendix A. By substituting Eqs. (A2) and (A3) for $\dot{\omega}_x^0$ and $\dot{\omega}_y^0$ into Eqs. (28) and (29), and performing the ensemble average $S_4^{\gamma} = \left\langle \ddot{u}_{\gamma}^2(0) \right\rangle$, we obtain the fourth moments, viz.

$$\begin{aligned}S_4^{\parallel} = \frac{1}{2\eta^4}\Bigg[ & -\sigma - \frac{\xi^4}{16\sigma^3} + \frac{\xi^2(7-\eta_\alpha^2/\tau^2)}{4\sigma^2} + \frac{2\xi^2 - 11 + 2\eta_\alpha^2/\tau^2}{4\sigma} + \frac{\eta_\alpha^2}{\tau^2} - 1 \\ & + \frac{\left(\sigma - \frac{\xi^2}{4\sigma} + \frac{11}{2} - \frac{\eta_\alpha^2}{\tau^2}\right)\cosh\xi - \frac{\xi}{2}\left(1 - \frac{\xi^2}{4\sigma^2} + \frac{13}{2\sigma} - \frac{\eta_\alpha^2}{\sigma\tau^2}\right)\sinh\xi}{B(\xi,\sigma)}\Bigg],\end{aligned} \tag{30}$$

$$\begin{aligned}S_4^{\perp} = \frac{1}{4\eta^4}\Bigg[ & \frac{\xi^4}{16\sigma^3} - \frac{\xi^2(7-\eta_\alpha^2/\tau^2)}{4\sigma^2} - \frac{\xi^2 - 11 + 2\eta_\alpha^2/\tau^2}{4\sigma} + \frac{\eta_\alpha^2}{\tau^2} + \frac{7}{2} \\ & + \frac{\left(2\sigma + \frac{\xi^2}{4\sigma} - \frac{11}{2} + \frac{\eta_\alpha^2}{\tau^2}\right)\cosh\xi - \frac{\xi}{2}\left(\frac{\xi^2}{4\sigma^2} - \frac{13}{2\sigma} + \frac{\eta_\alpha^2}{\sigma\tau^2}\right)\sinh\xi}{B(\xi,\sigma)}\Bigg],\end{aligned} \tag{31}$$

where $\eta_\alpha^2 = (1+\alpha^2)\eta^2$.

Higher-order moments can be obtained by analogy with the previous derivations. Due to the length of the calculation, only $S_6^{\gamma}$ and $S_8^{\parallel}$ for the undamped ($\alpha = 0$) rotating magnetization are given in Appendix B. Moreover, for $S_8^{\parallel}$ we also put $\sigma = 0$.

## 5. Results and discussion

The moments presented are not only a tool for estimating the correlation functions, but have important physical meaning. In accordance with the Kramers-Kronig relation the moment $S_0^g$ gives the static susceptibility $\chi'_g(0) = S_0^g$. For the physical interpretation of higher-order moments $S_{2n}^g$ we consider the *spectral* moments defined as [26,28]

$$M_{2n}^g = \lim_{\tau \to \infty} \frac{2}{\pi} \int_0^{\infty} \omega^{2n-1} \chi''_g(\omega + i/\tau)\, d\omega, \tag{32}$$

where $-\chi''_g(\omega)$ is the imaginary part of the susceptibility given by Eq. (20). These spectral moments $M_{2n}^g$ coincide with the moments $S_{2n}^g$ under certain conditions. By substituting Eq. (20) into Eq. (32) we have $M_2^g = S_2^g$, while $M_{2n}^g$ is not finite for $n > 1$. We note that all odd spectral moments $M_{2n-1}^g$ are vanishing. However, if $S_2^g = 0$, then $M_4^g = S_4^g$ and all $M_{2n}^g$ are not finite for $n > 2$, etc. Hence, it follows that the absorption coefficient $A_g(\omega) \sim \omega \chi''_g(\omega)$ decreases with increasing frequency as $A_g(\omega) \sim \omega^{-2n}$ if $M_{2n}^g$ is finite. Equation (32) for $n = 1$ is the Gordon sum rule, which determines the finite value of the integral absorption (the area under the absorption curve $A_g(\omega)$). Gordon's sum rule plays an important role in the theory of dielectrics. This rule as well as the spectral moments is a useful tool for extracting molecular parameters from the analysis of measured molecular spectra [41,42]. The same idea can be realized in the case of magnetic media. For example, Eqs. (25), (26) and (32) for $n = 1$ allow one to estimate the free energy parameters, as well as the inertial parameter, by analyzing the total absorption in a magnetic medium under various conditions.

In the non-inertial case the static susceptibility is not changed. However, the odd moments $S_{2n+1}^g$ are not vanishing now. For example, the first-order moments $S_1^g = -i\left\langle u_g(0)\dot{u}_g(0)\right\rangle$ are

$$S_1^{\parallel} = -\frac{i}{2\tau_N}\left(\frac{3\xi^2}{4\sigma^2} - 1 - \frac{3}{2\sigma} + \frac{3\left(\cosh\xi - \frac{\xi}{2\sigma}\sinh\xi\right)}{B(\xi,\sigma)}\right), \tag{33}$$

$$S_1^{\perp} = \frac{i}{4\tau_N}\left(\frac{3\xi^2}{4\sigma^2} - 1 - \frac{3}{2\sigma} + \frac{3\left(\cosh\xi - \frac{\xi}{2\sigma}\sinh\xi\right)}{B(\xi,\sigma)}\right), \qquad (34)$$

where $\tau_N$ is the free diffusion time

$$\tau_N = \frac{1+\alpha^2}{\alpha}\frac{\eta^2}{\tau} = \frac{1+\alpha^2}{\alpha}\frac{v\mu_0 M_S}{2\gamma kT}. \qquad (35)$$

We have from Eq. (20) for the non-inertial case

$$\chi_\gamma(\omega) = i\dot{C}_\gamma(0)/\omega + O(\omega^{-2}). \qquad (36)$$

Hence, the absorption coefficient $A_\gamma(\omega) \sim \omega\chi''_\gamma(\omega)$ in the limit $\omega \to \infty$ tends to a constant and the area under the absorption curve is infinite in the non-inertial case. This is not the case for the inertial magnetization dynamics ($S_1^\gamma = 0$), where the absorption coefficient tends to zero at high frequencies leading to the correct physical result, namely, to a finite area under the absorption curve (see Eq. (32) for $n = 1$). This phenomenon is well known in the theory of dielectrics, where considering the inertia of polar molecules ensures the transparency of the medium at high frequencies. In the limit $\alpha \to 0$ the first-order moments vanish $S_1^\gamma \to 0$.

The non-inertial analogs of $S_2^\gamma$ (compare with Eqs. (25) and (26)) are

$$S_2^{\parallel} = \frac{1}{4\tau_N^2}\left(-6 - \frac{\xi^4}{8\sigma^3} + \frac{9\xi^2}{2\sigma^2} - \frac{15}{2\sigma} + \frac{\xi^2}{\sigma} - 2\sigma + \frac{\left(15 - \frac{\xi^2}{2\sigma} + 2\sigma\right)\cosh\xi - \xi\left(1 - \frac{\xi^2}{4\sigma^2} + \frac{17}{2\sigma}\right)\sinh\xi}{B(\xi,\sigma)}\right), \qquad (37)$$

$$S_2^{\perp} = \frac{1}{8\tau_N^2}\left(3 - \frac{3}{\alpha^2} + \frac{\xi^4}{8\sigma^3} - \frac{9\xi^2}{2\sigma^2} - \frac{\xi^2 - 15}{2\sigma} + \frac{1}{\alpha^2}\left(\frac{\xi^2}{2\sigma} - 2\sigma\right) + \frac{\left(-15 + \frac{\xi^2}{2\sigma} + 4\sigma + \frac{6\sigma}{\alpha^2}\right)\cosh\xi - \xi\left(\frac{1}{\alpha^2} + \frac{\xi^2}{4\sigma^2} - \frac{17}{2\sigma}\right)\sinh\xi}{B(\xi,\sigma)}\right). \qquad (38)$$

In the limit $\alpha \to 0$ the moment $S_2^{\parallel}$ vanishes, but the moment $S_2^{\perp}$ does not.

As we noted, the inertia does not affect the static susceptibility $S_0^\gamma$. On the contrary all moments $S_{2n}^\gamma$ with $n \geq 1$ depend on the inertial relaxation time $\tau$ directly. Thus, the inertial relaxation time $\tau$ affects the area under the absorption curve $S_2^\gamma$ via the scaling factor $\eta$. Moreover, all moments $S_{2n}^\gamma$ with $n \geq 2$ depend on the damping parameter $\alpha$.

In the limit $\xi \to 0$ and $\sigma \to 0$, the moments tend to the isotropic case

$$S_0^{\parallel} = S_0^{\perp} = 1/3, \tag{39}$$

$$S_2^{\parallel} = S_2^{\perp} = 1/(3\eta^2), \tag{40}$$

and

$$S_4^{\parallel} = S_4^{\perp} = \left(2 + \eta_\alpha^2/\tau^2\right)/(3\eta^4). \tag{41}$$

The expressions for moments in the cases of a negligibly weak external field ($\sigma \gg \xi$) (such that the external field can be neglected) and in the opposite case of a very strong external field ($\xi \gg \sigma$) (such that the internal anisotropic potential can be neglected) are presented in Appendix C. In the latter case, the moments coincide with the moments derived in Ref. [43] for a symmetric top molecule, which has an electric dipole moment $\mu$ directed along the axis of symmetry of the top and rotates in an external uniform electric field $E$ after the following substitution

$$\eta^2 \to I/(2kT), \quad \eta^2/\tau^2 \to I_z/2I, \quad \xi \to \mu E/(kT), \tag{42}$$

where $I$ and $I_z$ are principal moments of inertia.

The correlation functions can be calculated directly using Eqs. (7)-(12), but this requires numerical solution of Eq. (1) and advanced numerical integration methods since the integrals in Eq. (9) are averages over all possible initial conditions. On the other hand, the correlation functions are well approximated by Eq. (17) at times shorter than $t/\eta < 1$. A comparison between numerical calculations and approximations of the correlation functions corresponding to inertial undamped ($\alpha \to 0$) magnetization dynamics in a strong uniform external field ($\xi \gg \sigma$) is shown in Figs. 2 and 3. Thus, the moments can be used as a fairly simple and satisfactory estimation of the numerical integration of the correlation functions at short times, thereby enabling the capture of fast oscillations. These fast oscillations are caused by the inertial term in the ILLG equation and are absent in the non-inertial case. A similar picture can be seen in the magnetization trajectories, i.e. the ILLG equation predicts fast nutational oscillations accompanying precessional motion of the magnetization, while the LLG equation describes precessional motion only [44].

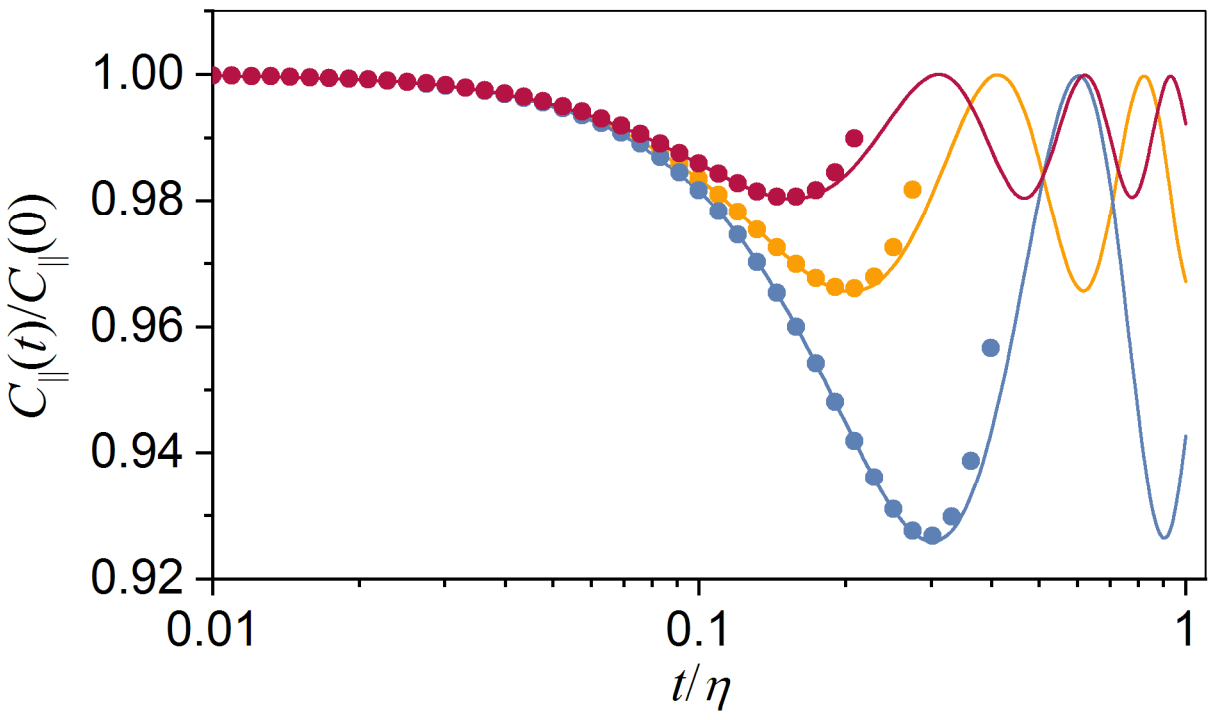

FIG. 2. Longitudinal correlation function $C_{\|}(t)$ for the field parameter $\xi = 5$ ($\xi \gg \sigma$ and $\alpha \to 0$) and various values of the inertial relaxation time, namely: (blue) $\eta/\tau = 10$, (orange) $\eta/\tau = 15$, and (red) $\eta/\tau = 20$. Solid lines: numerical calculations Eqs. (1), (7), (9)-(12); symbols: series approximation Eq. (17) with $n_{\max} = 4$.

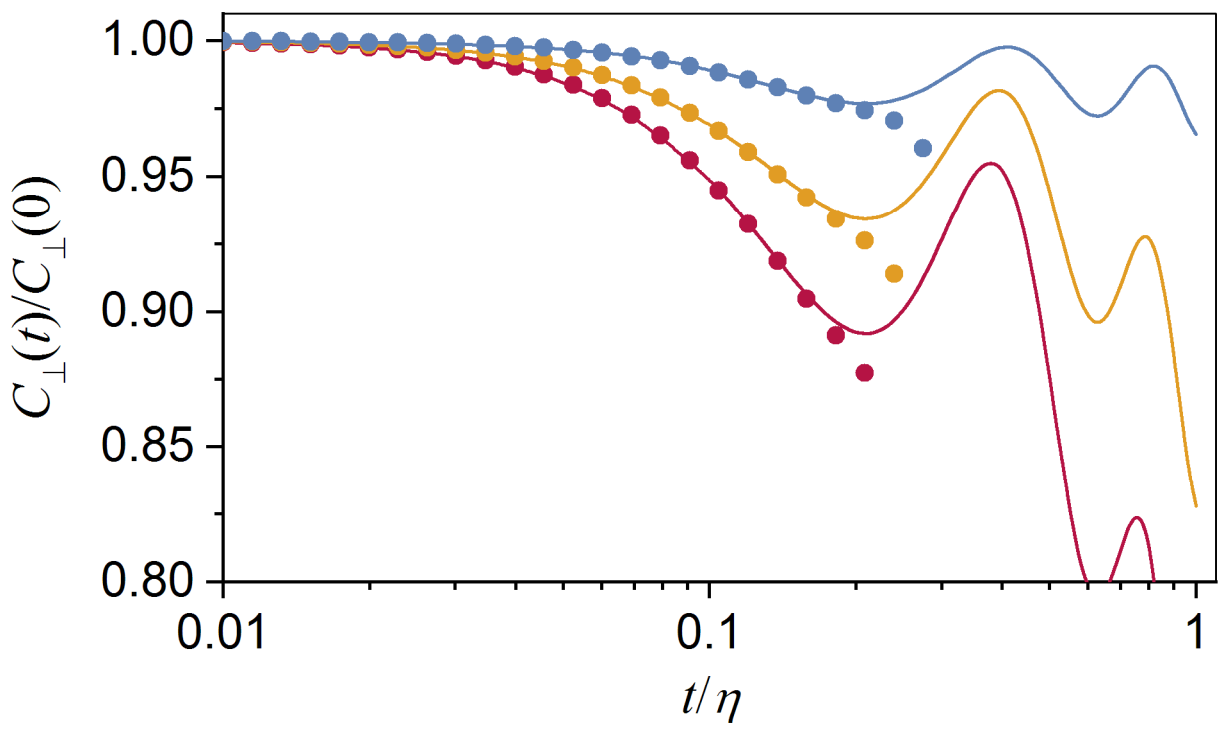

FIG. 3. Transverse correlation function $C_{\perp}(t)$ for the inertial relaxation time $\eta/\tau = 15$, $\alpha \to 0$ and various values of the field parameter $\xi \gg \sigma$, namely: (blue) $\xi = 5$, (orange) $\xi = 15$, and (red) $\xi = 25$. Solid lines: numerical calculations Eqs. (1), (8), (9)-(12); symbols: series approximation Eq. (17) with $n_{\max} = 3$.

### 6. Conclusions

We have presented the expressions for the moments up to the sixth order for the case of a cyclic symmetry of the free energy of ferromagnetic nanoparticles. Magnetic particles with uniaxial anisotropy in an external field directed perpendicular to the easy axis [31], with biaxial [31,45,46] and cubic anisotropies [47–50] can be investigated using the method presented. However, if circular symmetry of the magnetocrystalline anisotropy is broken, the moments can only be found in quadrature, and the expressions are quite complicated. In this case, it is preferable to calculate the moments numerically.

Because of the difficulties in extracting these moments from experimental spectral band shapes, it is unlikely that the complete expressions for moments higher than the fourth order will be directly useful. However, knowledge of the second and fourth moments are useful for investigation of the static susceptibility, area under the absorption curve, and extraction of the potential parameters from the spectra.

The method presented can also be employed to describe the relaxation dynamics in the processes of magnetization or magnetization reversals. These dynamics correspond to a process of reaching thermodynamic equilibrium, and this process is described by correlation functions. For example, the variation of the external field from $\mathbf{H}+\Delta\mathbf{H}$ to $\mathbf{H}$ ($|\Delta\mathbf{H}| = |\mathbf{H}|$) results in a disturbance of the equilibrium state of the spin system and the system relaxes to a new equilibrium state. Magnetic relaxation determines the linewidth of ferromagnetic, antiferromagnetic, and nutation resonances. Bloch [31] introduced the phenomenological approach that the evolution of the magnetization towards equilibrium was exponential, however embodying two distinct time constants: $T_{\parallel}$ for the longitudinal component and $T_{\perp}$ for the transverse component. There is an intuitive connection between the time evolution of correlation functions and the time evolution of macroscopic systems: on average, the correlation function evolves in time in the same manner. The regression of microscopic thermal fluctuations at equilibrium follows the macroscopic law of relaxation of small nonequilibrium disturbances. As the values of microscopic variables separated by large timescales, should be uncorrelated, the evolution in time of a correlation function can be viewed from a physical standpoint as the system gradually forgetting the initial conditions. This concept is reflected in the van Vleck–Weisskopf model, in which the longitudinal and transverse correlation functions are expressed as [25]

$$C_{\parallel}(t) = C_{\parallel}^{un}(t)\exp\left(-\frac{t}{T_{\parallel}}\right), \quad C_{\perp}(t) = C_{\perp}^{un}(t)\exp\left(-\frac{t}{T_{\perp}}\right), \tag{43}$$

where the correlation functions $C_{\parallel}^{un}(t)$ and $C_{\perp}^{un}(t)$ correspond to the undamped dynamics of magnetization. The model is commonly used to describe orientational relaxation in gases and liquids by considering inertial effects [25,51]. For the correlation functions given by Eq. (43) the moment $S_1^{\gamma}$ is not vanishing, viz.,

$$S_1^{\gamma} = i C_{\gamma}^{un}(0) / T_{\gamma} . \qquad (44)$$

Hence the area under the absorption curve $\omega\chi''_{\gamma}(\omega)$ is infinite for the van Vleck–Weisskopf model. The shortcomings of the van Vleck–Weisskopf model have been overcome in a more complex *J*-diffusion model [25,43]. For the *J*-diffusion model the first-order moment is vanishing.

Next application of the method of moments can be found in the calculation of the cumulants, which are used in statistical physics to describe complex distributions [52,53]. The cumulants $K_n^{\gamma}$ can be expressed via moments and vice versa. A very convenient form in terms of determinants reads [52]

$$K_n^{\gamma} = (-1)^{n-1} \det \begin{pmatrix} S_1^{\gamma} & 1 & 0 & 0 & \cdots \\ S_2^{\gamma} & S_1^{\gamma} & 1 & 0 & \cdots \\ S_3^{\gamma} & S_2^{\gamma} & \binom{2}{1} S_1^{\gamma} & 1 & \cdots \\ \vdots & \vdots & \vdots & \vdots & \ddots \\ S_n^{\gamma} & S_{n-1}^{\gamma} & \binom{n-1}{1} S_{n-2}^{\gamma} & \binom{n-1}{2} S_{n-3}^{\gamma} & \cdots \end{pmatrix},$$

where $\binom{n}{m} = \frac{n!}{(n-m)!m!}$ are the binominal coefficients. We anticipate that future work will demonstrate the application of the moments for cumulant calculation in magnetic materials, where magnetization demonstrates inertial dynamics. Moreover, the presented method can also be effectively used for non-Markovian systems, especially if stationary solutions of the master equation for the distribution function can be found in analytical form.

In summary, we note that it is difficult to evaluate all the moments, since the analytical expression for the moments become more complicated as one goes to higher-order moments. Thus, the correlation function and susceptibility are approximated by finite series, which are valid only at short times. Even if the time series converged at larger times, one would seek an alternative method of evaluation of the correlation function for long times. However, the moment method sharply reduces the complexity of correlation functions analysis. Indeed, the precise expressions of these functions in Eqs. (7) and (8) necessitate the solution of the ILLG equation followed by averaging the dynamic functions over the initial conditions. However, the moment method allows one to simplify this, i.e., one needs to find the average values of the correlation functions over coordinates and velocities at the initial times only. Moreover, the proposed method provides a basis to calculate the static susceptibility, integral reorientation time, and other characteristics of ferromagnetic nanosystems. It is possible to investigate the dependence of these characteristics on the free energy parameters of ferromagnetic nanoparticles.

## Acknowledgments

We thank Y.P. Kalmykov for useful comments and suggestions. S.T. and A.T. express their acknowledgment to the Foundation for the Advancement of Theoretical Physics and Mathematics “BASIS” (Grant ID 22-1-1-28-1).

## Appendix A: Time derivatives of the angular velocity components

The time derivatives of the angular velocity components, $\dot{\omega}_x$ and $\dot{\omega}_y$, can be determined from the ILLG equation (1). This equation is used since it provides the relation between the high-order derivatives of coordinates and velocities. To find $\dot{\omega}_x$ and $\dot{\omega}_y$ we express the vector **u** and its time derivatives occurring in Eq. (1) in spherical polar coordinates $\{e_r, e_\vartheta, e_\varphi\}$ as

$$\mathbf{u}=\begin{pmatrix}1\\0\\0\end{pmatrix},\quad \dot{\mathbf{u}}=\begin{pmatrix}0\\ \dot{\vartheta}\\ \dot{\varphi}\sin\vartheta\end{pmatrix},\quad \ddot{\mathbf{u}}=\begin{pmatrix}-\dot{\vartheta}^2-\dot{\varphi}^2\sin^2\vartheta\\ \ddot{\vartheta}-\dot{\varphi}^2\cos\vartheta\sin\vartheta\\ 2\dot{\vartheta}\dot{\varphi}\cos\vartheta+\ddot{\varphi}\sin\vartheta\end{pmatrix}. \tag{A1}$$

By substituting all these vectors into the vector Eq. (1), and cinsidering Eqs. (3) and (16), we derive two coupled equations expressing the initial values of the time derivatives of the angular velocity components $\dot{\omega}_x^0=\dot{\omega}_x(0)$ and $\dot{\omega}_y^0=\dot{\omega}_y(0)$ via initial values of the angular velocity components $\omega_x^0=\dot{\vartheta}(0), \omega_y^0=\dot{\varphi}(0)\sin\vartheta(0)$ and angles $\vartheta_0, \varphi_0$ [19]

$$\dot{\omega}_x^0=\frac{1}{\tau}\omega_y^0+\left(\omega_y^0\right)^2\cot\vartheta_0-\frac{\alpha}{\tau}\omega_x^0-\frac{1}{2\eta^2}\left(2\sigma\cos\vartheta_0+\xi\right)\sin\vartheta_0, \tag{A2}$$

$$\dot{\omega}_y^0=-\frac{1}{\tau}\omega_x^0-\omega_x^0\omega_y^0\cot\vartheta_0-\frac{\alpha}{\tau}\omega_y^0. \tag{A3}$$

## Appendix B: High order moments

For the calculation of the moments $S_6^{\parallel}$ and $S_6^{\perp}$, we need the time derivatives $\dddot{\omega}_{\parallel}(0)$ and $\dddot{\omega}_{\perp}(0)$. By using Eqs. (A2) and (A3), we obtain for $\dddot{\omega}_{\parallel}(0)$ and $\dddot{\omega}_{\perp}(0)$:

$$\dddot{\omega}_{\parallel}(0)=\frac{1}{\tau^2}\omega_x^0\sin\vartheta_0+\left(\left(\omega_x^0\right)^3+\omega_x^0\left(\omega_y^0\right)^2\right)\sin\vartheta_0+\frac{\xi}{\eta^2}\omega_x^0\sin 2\vartheta_0+\frac{\sigma}{\eta^2}\omega_x^0\sin\vartheta_0\left(5\cos^2\vartheta_0-1\right), \tag{B1}$$

$$\begin{aligned}\dddot{\omega}_{\perp}(0)=&\left(\frac{1}{\tau^2}+\left(\omega_x^0\right)^2+\left(\omega_y^0\right)^2\right)\left(\omega_y^0\sin\varphi_0-\omega_x^0\cos\vartheta_0\cos\varphi_0\right)\\&+\frac{\xi}{2\eta^2}\left(\omega_x^0(4\sin^2\vartheta_0-1)\cos\varphi_0+\omega_y^0\cos\vartheta_0\sin\varphi_0-\frac{1}{\tau}\sin\vartheta_0\sin\varphi_0\right)\\&-\frac{1}{\tau}\frac{\sigma}{\eta^2}\cos\vartheta_0\sin\vartheta_0\sin\varphi_0+\frac{\sigma}{\eta^2}\omega_x^0(5\sin^2\vartheta_0-1)\cos\vartheta_0\cos\varphi_0+\frac{\sigma}{\eta^2}\omega_y^0\cos^2\vartheta_0\sin\varphi_0.\end{aligned} \tag{B2}$$

By using Eqs. (B1) and (B2) and performing the ensemble averaging, we obtain the moments $S_6^{\parallel,\perp}=\left\langle\dddot{\omega}^2_{\parallel,\perp}(0)\right\rangle$, viz.,

$$S_6^{\parallel,\perp}=\eta^{-6}\left[A^{\parallel,\perp}+\frac{A_c^{\parallel,\perp}\cosh\xi+A_s^{\parallel,\perp}\sinh\xi}{B(\xi,\sigma)}\right], \tag{B3}$$

where

$$\begin{aligned}A^{\perp}=&-\frac{69}{16}+\frac{\xi^2}{16}+\frac{\xi^6}{256\sigma^4}-\frac{63\xi^4}{128\sigma^3}+\frac{453\xi^2}{64\sigma^2}-\frac{\xi^4}{32\sigma^2}-\frac{279}{32\sigma}+\frac{69\xi^2}{32\sigma}-2\sigma\\&+\frac{\eta^4}{4\tau^4}+\frac{\eta^4\xi^2}{16\sigma^2\tau^4}-\frac{\eta^4}{8\sigma\tau^4}+\frac{11\eta^2}{8\tau^2}+\frac{\eta^2\xi^4}{32\sigma^3\tau^2}-\frac{7\eta^2\xi^2}{8\sigma^2\tau^2}+\frac{11\eta^2}{8\sigma\tau^2}-\frac{\eta^2\xi^2}{16\sigma\tau^2}-\frac{\eta^2\sigma}{4\tau^2},\\A_c^{\perp}=&\frac{279}{16}-\frac{\xi^2}{16}+\frac{\xi^4}{64\sigma^2}-\frac{15\xi^2}{8\sigma}+\sigma+\frac{\sigma^2}{2}+\frac{\eta^4}{4\tau^4}-\frac{11\eta^2}{4\tau^2}+\frac{\eta^2\xi^2}{8\sigma\tau^2}+\frac{7\eta^2\sigma}{4\tau^2},\\A_s^{\perp}=&-\frac{5\xi}{8}-\frac{\xi^5}{128\sigma^3}+\frac{31\xi^3}{32\sigma^2}-\frac{393\xi}{32\sigma}+\frac{\xi^3}{32\sigma}+\frac{\xi\sigma}{4}-\frac{\eta^4\xi}{8\sigma\tau^4}-\frac{\eta^2\xi}{8\tau^2}-\frac{\eta^2\xi^3}{16\sigma^2\tau^2}+\frac{13\eta^2\xi}{8\sigma\tau^2},\end{aligned} \tag{B4}$$

$$\begin{aligned}A^{\parallel}=&\frac{\xi^4+2\xi^2(3-4\sigma)\sigma+8\sigma^2\left(6-13\sigma+2\sigma^2\right)}{32\sigma^2}\\&+\frac{\eta^4}{\tau^4}+\frac{\eta^2\left(3\xi^2+2(7-6\sigma)\sigma\right)}{8\sigma\tau^2}-2A^{\perp},\\A_c^{\parallel}=&\frac{\xi^2}{8}+9\sigma+\frac{\sigma^2}{2}+\frac{9\eta^2\sigma}{2\tau^2}-2A_c^{\perp},\\A_s^{\parallel}=&-\frac{\xi}{2}-\frac{\xi^3}{16\sigma}+\frac{3\xi\sigma}{4}-\frac{3\eta^2\xi}{4\tau^2}-2A_s^{\perp}.\end{aligned} \tag{B5}$$

In the isotropic case ($\sigma\to 0$ and $\xi\to 0$) Eqs. (C8) and (C9) are simplified to yield

$$S_6^{\parallel}=S_6^{\perp}=\left(\eta^4/\tau^4+4\eta^2/\tau^2+6\right)/(3\eta^6). \tag{B6}$$

Here we also present the analytical expression for the moment $S_8^{\parallel} = \left\langle \left( u_{\parallel}^{(4)}(0) \right)^2 \right\rangle$ for the case $\sigma = 0$, which is used in the calculations in Fig. 2. The moment $S_8^{\parallel}$ can be obtained by analogy with the derivation of the moment $S_6^{\parallel}$ by noting that

$$\begin{aligned} u_{\parallel}^{(4)}(0) &= \left( \frac{1}{\tau^2} + \omega_x^2 + \omega_y^2 \right) \left( \dot{\omega}_x \sin\vartheta + \omega_x^2 \cos\vartheta \right) \\ &+ 2\omega_x \left( \omega_x \dot{\omega}_x + \omega_y \dot{\omega}_y \right) \sin\vartheta + \frac{\xi}{\eta^2} \left( \dot{\omega}_x \sin 2\vartheta + 2\omega_x^2 \cos 2\vartheta \right). \end{aligned} \tag{B7}$$

Thus, by using Eqs. (A2) and (A3), substituting Eq. (B7) into Eq. (18), and performing the ensemble averaging, we obtain

$$\begin{aligned} S_8^{\parallel} = \frac{1}{\eta^8} \Bigg\{ & 8\xi^2 + 154 - 24\frac{\eta^2}{\tau^2} + 6\frac{\eta^4}{\tau^4} \\ & + \left( 4\xi^2 \left( 3\frac{\eta^2}{\tau^2} - 8 \right) - 438 + 90\frac{\eta^2}{\tau^2} - 12\frac{\eta^4}{\tau^4} + \frac{\eta^6}{\tau^6} \right) \xi^{-1} \left( \coth\xi - \xi^{-1} \right) \Bigg\}. \end{aligned} \tag{B8}$$

Analytical expressions for the moments of higher orders are quite complex, and their derivation is quite laborious. However, they can still be calculated numerically without much difficulty.

## Appendix C: Weak ($\sigma \gg \xi$) and strong ($\xi \gg \sigma$) external fields

If magnetic particles are placed in a very strong external field ($\xi \gg \sigma$) (such that the internal anisotropic potential can be neglected), the free energy in Eq. (3) becomes

$$V(\vartheta)/(kT) = -\xi \cos\vartheta . \tag{C1}$$

The moments reduce to the following equations

$$S_0^{\parallel} = 1 - \coth^2\xi + \xi^{-2}, \tag{C2}$$

$$S_0^{\perp} = \xi^{-1}\coth\xi - \xi^{-2}, \tag{C3}$$

$$S_2^{\parallel} = \frac{\xi\coth\xi - 1}{\eta^2\xi^2}, \tag{C4}$$

$$S_2^{\perp} = \frac{\xi^2 - \xi\coth\xi + 1}{2\eta^2\xi^2}, \tag{C5}$$

$$S_4^{\parallel} = \frac{1}{\eta^4} \left[ 2 - \frac{1}{\xi} \left( 4 - \frac{\eta_a^2}{\tau^2} \right) \left( \coth\xi - \frac{1}{\xi} \right) \right], \tag{C6}$$

$$S_4^{\perp} = \frac{1}{2\eta^4}\left[\frac{\eta_a^2}{\tau^2} + \left(\frac{\xi}{2} + \frac{1}{\xi}\left(4 - \frac{\eta_a^2}{\tau^2}\right)\right)\left(\coth\xi - \frac{1}{\xi}\right)\right], \tag{C7}$$

$$S_6^{\parallel} = \frac{1}{\eta^6}\left\{-8 + 4\frac{\eta^2}{\tau^2} + \left(30 + 4\xi^2 - 8\frac{\eta^2}{\tau^2} + \frac{\eta^4}{\tau^4}\right)\xi^{-1}\left(\coth\xi - \xi^{-1}\right)\right\}, \tag{C8}$$

$$S_6^{\perp} = \frac{1}{8\eta^6}\left\{56 + \xi^2 + 4\frac{\eta^4}{\tau^4} - \left(120 - 32\frac{\eta^2}{\tau^2} + 4\frac{\eta^4}{\tau^4} + \xi^2\left(1 - 6\frac{\eta^2}{\tau^2}\right)\right)\xi^{-1}\left(\coth\xi - \xi^{-1}\right)\right\}. \tag{C9}$$

In the negligibly weak external field ($\xi \to 0$), the free energy in Eq. (3) becomes

$$V(\vartheta)/(kT) = -\sigma\cos^2\vartheta . \tag{C10}$$

The equations for the moments simplify to yield

$$S_0^{\parallel} = \frac{1}{2}\left(\frac{1}{\sqrt{\sigma}\,F_D(\sqrt{\sigma})} - \frac{1}{\sigma}\right), \tag{C11}$$

$$S_0^{\perp} = \frac{1}{4}\left(2 + \frac{1}{\sigma} - \frac{1}{\sqrt{\sigma}\,F_D(\sqrt{\sigma})}\right), \tag{C12}$$

$$S_2^{\parallel} = \frac{1}{4\eta^2}\left(2 + \frac{1}{\sigma} - \frac{1}{\sqrt{\sigma}\,F_D(\sqrt{\sigma})}\right), \tag{C13}$$

$$S_2^{\perp} = \frac{1}{8\eta^2}\left(2 - \frac{1}{\sigma} + \frac{1}{\sqrt{\sigma}\,F_D(\sqrt{\sigma})}\right), \tag{C14}$$

$$S_4^{\parallel} = \frac{1}{8\eta^4}\left(-4 - 4\sigma - \frac{11}{\sigma} + \frac{\eta_a^2}{\tau^2}\left(4 + \frac{2}{\sigma}\right) + \frac{11 + 2\sigma - 2\eta_a^2/\tau^2}{\sqrt{\sigma}\,F_D(\sqrt{\sigma})}\right), \tag{C15}$$

$$S_4^{\perp} = \frac{1}{16\eta^4}\left(14 + \frac{11}{\sigma} + \frac{\eta_a^2}{\tau^2}\left(4 - \frac{2}{\sigma}\right) + \frac{4\sigma - 11 + 2\eta_a^2/\tau^2}{\sqrt{\sigma}\,F_D(\sqrt{\sigma})}\right), \tag{C16}$$

$$S_6^{\parallel} = \frac{1}{16\eta^6}\left[162 + \frac{279}{\sigma} + 12\sigma + 8\sigma^2 + \frac{8\eta^4}{\tau^4}\left(1 + \frac{1}{2\sigma}\right) - \frac{16\eta^2}{\tau^2}\left(1 + \sigma + \frac{11^2}{4\sigma}\right) + \frac{-279 + 56\sigma - 4\sigma^2 + (44 + 8\sigma)(\eta/\tau)^2 - 4(\eta/\tau)^4}{\sqrt{\sigma}\,F_D\left(\sqrt{\sigma}\right)}\right], \tag{C17}$$

$$S_6^{\wedge} = \frac{1}{32\eta^6}\left[-138 - \frac{279}{\sigma} - 64\sigma - \frac{4\eta^2}{\tau^2}\left(-11+2\sigma-\frac{11}{\sigma}\right) + \frac{8\eta^4}{\tau^4}\left(1-\frac{1}{2\sigma}\right) + \frac{279+16\sigma+8\sigma^2+4\left(-11+7\sigma\right)(\eta/\tau)^2+4(\eta/\tau)^4}{\sqrt{\sigma}F_D\left(\sqrt{\sigma}\right)}\right]. \tag{C18}$$